\documentclass[aps,prb,twocolumn,showpacs,amssymb]{revtex4}

\usepackage[dvips]{graphicx}
\usepackage{dcolumn}
\usepackage{bm}
\def\en{\end{equation}}

\def\bea{\begin{eqnarray}}
\def\ena{\end{eqnarray}}

\begin{document}

\title{Spatiotemporal heterogeneity of local free volumes in highly supercooled liquid}

\author{Hayato Shiba}
\email{shiba@issp.u-tokyo.ac.jp}
\affiliation{Institute for Solid State Physics, University of Tokyo, Chiba 277-8581, Japan}
\author{Takeshi Kawasaki}
\affiliation{Department of Physics, Kyoto University, Kyoto 606-8502, Japan}
\altaffiliation{Present address: Laboratoire Charles Coulomb, UMR 5221, CNRS and 
Universit\'e Montpellier 2, 34095 Montpellier, France}

\date{\today}

\begin{abstract}
We discuss the spatiotemporal behavior of local density 
and its relation to dynamical heterogeneity in a highly supercooled liquid
by using molecular dynamics simulations of a binary mixture with different 
particle sizes in two dimensions. To trace voids   heterogeneously existing with lower local densities, 
which move along with the structural relaxation, we employ the minimum local density for each particle 
in a time window whose width is set along with the structural relaxation time.
Particles subject to free volumes correspond well to the configuration rearranging region of dynamical heterogeneity.
While the correlation length for dynamical heterogeneity grows with temperature decrease, no growth
in the correlation length of heterogeneity in the minimum local density distribution takes place. 
A comparison of these results with those of  normal mode analysis reveals
that superpositions of lower-frequency soft modes extending over the free volumes exhibit spatial
correlation with the broken bonds. This observation 
suggests a possibility that long-ranged vibration modes 
facilitate the interactions between fragile regions represented by free volumes,
to induce dynamical correlations at a large scale.
\end{abstract}

\maketitle

\section{Introduction}
The reason for the drastic slowdown in dynamics when liquids 
are cooled toward the glass transition temperature
has been a long-standing problem.
In the last two decades, numerous 
efforts have been made to study this problem via molecular dynamics (MD) 
simulations.\cite{BinderKob}
The recently proposed  concept of 
``dynamic heterogeneity'' \cite{97YO_JPSJ,98YO_PRE,95Muranaka,97Kob,98Donati,
06BBMR,12SKO,13KawasakiJCP,2013Kim} has 
indicated that there could be a contrast between the
mobile and immobile regions of supercooled liquids on a large scale,
as if there were critical fluctuations hidden behind the disordered configuration. 
Over the past several decades, suggestions have been made 
regarding the static origin of the cooperative rearranging 
region.\cite{AG,RFOT} 
In recent literature, considerable attention 
has been focused on the relation of dynamical heterogeneity to the 
structural heterogeneity of medium length-scale crystalline 
order,\cite{07Kawasaki,10TanakaNMat}
icosahedral order, \cite{02Dzugutov,2012Leocmach} 
local potential energy, \cite{06Poole} and so on.
There are also continuing discussions about whether 
or not growing length scales of structures
occur in the vitrified states.\cite{13TorquatoJCP,12Coslovich}

Density fluctuation is one of the candidate origins for such a static entity. 
In the context of certain experiments, particularly on metallic, polymer,
and colloidal glasses, density fluctuations have been observed over
small-to-large length scales.\cite{91Fischer,91vanMegen,93Fischer,94Kanaya,00Fischer}
One of the primary theoretical challenges thus far has 
been the construction of
statistical descriptions of solids and glasses including the effects of 
vacancies or local free volume. \cite{59Turnbull,76Cohen,79Argon,
98Ediger,98Falk,02Lemaitre,04OnukiPRAMANA}
In MD simulations of binary mixtures usually employed in stuies 
of supercooled liquids, density fluctuations are too weak 
to be captured clearly, \cite{98YO_PRE,06WCH,10TanakaNMat}
and the correlation of such fluctuations
to dynamical heterogeneity has only partially been observed.\cite{06Ladadwa}
In crystals subject tohomogeneous melting, wherein the density fluctuation is 
more pronounced than in glass, diffusion of interstitial defects 
with relatively smaller local densities is ascribed as the cause 
of cooperative motion in two\cite{09ShibaEPL} and 
three\cite{05Granato,13DouglasJCP} dimensions. 
By considering how similar the dynamical properties
are in ordered and disordered systems, 
the role of density fluctuation in dynamical heterogeneity can 
be revealed much more clearly.

In spite of  difficulties in studying the local density fluctuations, 
one method has recently enabled the possibility of capturing 
the ``inherent structure'' as a localized soft mode, which 
can be calculated from a static configuration of particles
in a supercooled liquid with the use of normal mode analysis.\cite{08WC,12Matsuoka}
Since this approach is, to our knowledge, the only established approach to 
predict tendencies of heterogeneity of dynamic propensity
from a momentary configuration of the particles, 
it can be the bridge between dynamical heterogeneity and its origin. 
In this study, we examine the heterogeneities of the local density distributions 
and dynamics by capturing and tracing weakly existing density 
fluctuations in a constructive manner, 
and subsequently, we investigate its relationship to the inherent 
structure to clarify the probable role of 
the density fluctuation in dynamical heterogeneity.
The composition of this paper is as follows:
In Sec. \ref{sec:method}, the simulation model and methods are provided.
The numerical analysis of the local density fluctuation
and its heterogeneity is described in Secs. \ref{sec:res}
A-C. In Sec. \ref{sec:res} D, the result is compared with 
the heterogeneity of normal modes. In Sec. \ref{sec:conclusion}
concludes the paper. 

\section{Numerical method} \label{sec:method}
We investigate a two-dimensional (2D) binary mixture 
composed of two atomic species, 1 and 2, with 
$N_1=N_2=32000$ particles. 
The particles interact via the soft-core potentials 
$v_{\alpha\beta} (r) = \epsilon (\sigma_{\alpha\beta} /r)^{12}+C_{\alpha\beta}$
where 
 $\sigma_{\alpha\beta} = (\sigma_\alpha + \sigma_\beta)/2$  and $r$ denote
 the interaction lengths and the distance between two particles respectively,
with $\alpha,\beta\in \{1,2\}$.
The interaction is truncated at $r= 4.5\sigma_1$ and 
a constant $C_{\alpha\beta}$ is set so as to ensure continuity of the potential
at the cutoff.
The size ratio between the two species is $\sigma_2/\sigma_1 =1.4$
to prevent crystallization, and the mass is set as $m_2/m_1 = (\sigma_2/\sigma_1)^2$. 
The particle density is fixed at a high value of 
$\phi = (N_1\sigma_1^2 + N_2\sigma_2^2)/V = 1.2$,
and therefore the particle configurations are jammed in the supercooled state. 
No tendency of phase separation is detected in our computation times. 
Space, time, and temperature are measured in units of $\sigma_1$, 
$\tau_0 = \sqrt{m_1\sigma_1^2 /\epsilon}$, and $\epsilon/k_B$, respectively.
The dependence of the $\alpha$-relaxation time
on the temperature $T$ is shown in Table \ref{tab:relaxtime}. 
A sufficiently long annealing time $t_A$ is chosen
($t_A > 30\tau_\alpha$)
with the time step being $\Delta t=0.005$. 
The aging effect was negligible in the course of 
calculation of pressure, density time correlations, etc.
After performing the above procedure at each temperature,
we begin to collect the data. 
This time point in each of our simulations is denoted as $t=0$ in 
the following. 

\begin{table*}
\caption{\label{tab:relaxtime} Dependence of $\alpha$-relaxation time ($\tau_\alpha$) and bond-relation time ($\tau_b$) on $T$.} 
\begin{ruledtabular}
\begin{tabular}{llllllp{2in}}
$T$ & 0.56 & 0.64 & 0.72 & 0.80 & 0.96 & 1.20 \\
\hline
$\tau_\alpha$ & $2.14\times 10^3$ & $75.5$ & 10.5 & 5.39 & 2.86 & 1.47 \\
$\tau_b$  & $1.41\times 10^5$ & $1.90\times 10^4$ & $5.25\times 10^3$ & $1.87\times 10^3$ & $5.35\times 10^2$ & $1.85\times 10^2$ \\ 
\end{tabular}
\end{ruledtabular} 
\end{table*}

\begin{figure}
\includegraphics[width=0.9\linewidth]{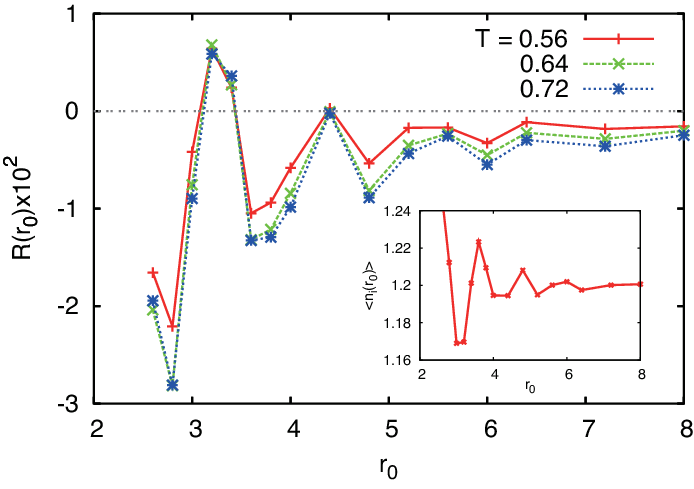}
\caption{\label{fig:LDdiff}
Degree of deviation $R(r_0)$ of the local densities 
 around the broken bonds $[ n_i(r_0) ]_{\rm b.b.}$
(obtained at intervals of $\Delta t=2\times 10^2$)
from that averaged over the entire system $[ n_i(r_0) ]_N$, 
plotted as a function of the averaging radius $r_0$. 
The data for $T=0.56, 0.64,$ and $0.72$ are shown. 
Inset: $\langle n_i (r_0)\rangle$ is shown for $T=0.56$. 
}
\end{figure}

\section{Results} \label{sec:res}
\subsection{Local density}
First, for each particle $i$, we define the local density 
by counting the particles within the distance $\Delta r
= |\bm{r}-\bm{r}_i | \le r_0$ from $i$
\begin{equation}
n_i(r_0)  = \frac{1}{\pi r_0^2}
\int_{\Delta r < r_0} \!\!\!\!\!\! d\bm{r}\
\sigma_j^2 \delta (\bm{r}_j -\bm{r} ).
\end{equation}
This local density is weighted by the assumed area $\sigma_j^2$. 
The inset of Fig. 1 shows the long-time average of the local density
$[\overline{n_i (r_0)} ]_N$ at $T=0.56$.
Here, $[ \mathcal{A}_i ]_N = N^{-1}\sum_i \mathcal{A}_i$ 
denotes the average over all the particles. 
This quantity corresponds to a radial
distribution function $g(r_0)$ defined in a naive manner.

\begin{figure}
\includegraphics[width=0.95\linewidth]{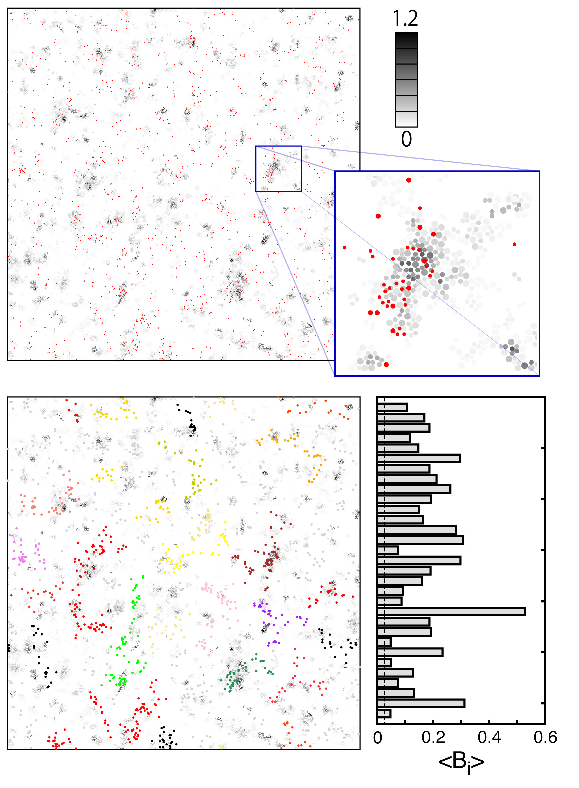}
\caption{\label{fig:FV_ST}
 (Top panel)
The red points show the distribution of 2\% of the particles having 
smaller local densities $\langle n_i(r_0)\rangle_{\rm IE}$, which is averaged over $t\in [0,10]$,
in an IE of 32 runs at $T=0.56$. 
The number distribution of broken bonds for the same IE 
$\langle \hat{\mathcal{B}}(\bm{r},\Delta t)\rangle_{\rm IE}$ with $t\in [0,122]$
is shown according to the scale bar on the right. 
For clarity, a magnified figure is shown to the right.
(Bottom panel)
In the left, for the same particle configuration as in the top panel,
the particles are classified into several groups according to how 
they are close to each other. Each group is colored by a different color.
To the right, the average value of the broken bond density 
$\langle \mathcal{B}_i\rangle_{\rm IE}$ over each cluster region is shown, 
where the dotted line indicates the average broken bond density 
$\mathcal{B}_0=0.0268$.
}
\end{figure}

For the same simulation run, at appropriate time intervals, 
we estimate the local density in the 
configuration rearranging regions of dynamical heterogeneity
as follows: 
bonds are defined at each time $t_0$ as 
particle pairs between $i$ and $j$ satisfying the condition\cite{97YO_JPSJ,98YO_PRE,12SKO}
\begin{equation}
r_{ij}(t_0) \le A_1\sigma_{\alpha\beta}  \label{eq:bonded}
\end{equation}
and after a time interval $\Delta t$ bonds are regarded to be broken if 
\begin{equation}
r_{ij} (t_0+\Delta t) > A_2\sigma_{\alpha\beta}, \label{eq:broken}
\end{equation}
where the cutoffs are set to $A_1 =1.15$ and $A_2 =1.6$. 
Further, in the remainder of this paper, broken bonds
in the time interval $[t,t+ \Delta t]$ 
are defined in the same way.
We obtain the local density profile only for particles that
have undergone bond breakage in a time interval 
of $\Delta t = 2\times 10^2$. More precisely, 
we check whether all the bonds 
existing at time $t=t_0$ 
are broken after a time interval of $\Delta t$, and for the particles 
having at least one broken bond the
long-time average of the  local density $n_i$ is calculated.
Their average local density profile is denoted as $[n_i(r_0)]_{\rm b.b.}$. 
The main graph in Fig. 1 shows the long-time average 
$\overline{R(r_0, \Delta t)}$ of the relative degree of deviation 
as given by
\begin{equation}
R(r_0, \Delta t) = \left( \frac{[n_i(r_0)]_{\rm b.b.} }{[n_i(r_0)]_N} -1 \right). \label{eq:ratio}
\end{equation}
While $R(r_0)$ exhibits an oscillating behavior at small values of $r_0$ mainly 
due to the mixed contributions from larger and smaller components,
its value is systematically lower than zero 
when $r_0$ is sufficiently large. 
Thus, around the broken bonds, {\it i.e.}, in the configuration rearranging regions, 
the degree of local density can be characterized with a sufficiently 
large value of $r_0$ at each particle.
We employ $r_0=6.0$ in our following discussions.

\begin{figure}
\includegraphics[width=0.9\linewidth]{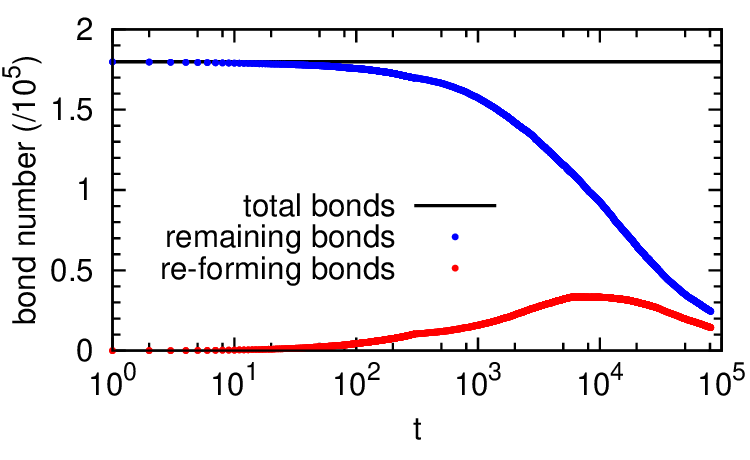}
\caption{\label{fig:bond}
Time development of the bond number for $T=0.64$.
The black line indicates the total number of the 
initial bonds, and the bonds are defined as pairs 
satisfying  $r_{ij}<A_1\sigma_{\alpha\beta}$ at $t=0$. 
The blue points indicate the number of 
remaining bonds at time $t$.
``re-forming bonds'', which denote
those bonds that have undergone bond
breakage and rebonding rebounded
in a time interval of $[0,t)$, 
are indicated by the red points.
}
\end{figure}

\begin{figure}
\includegraphics[width=0.75\linewidth]{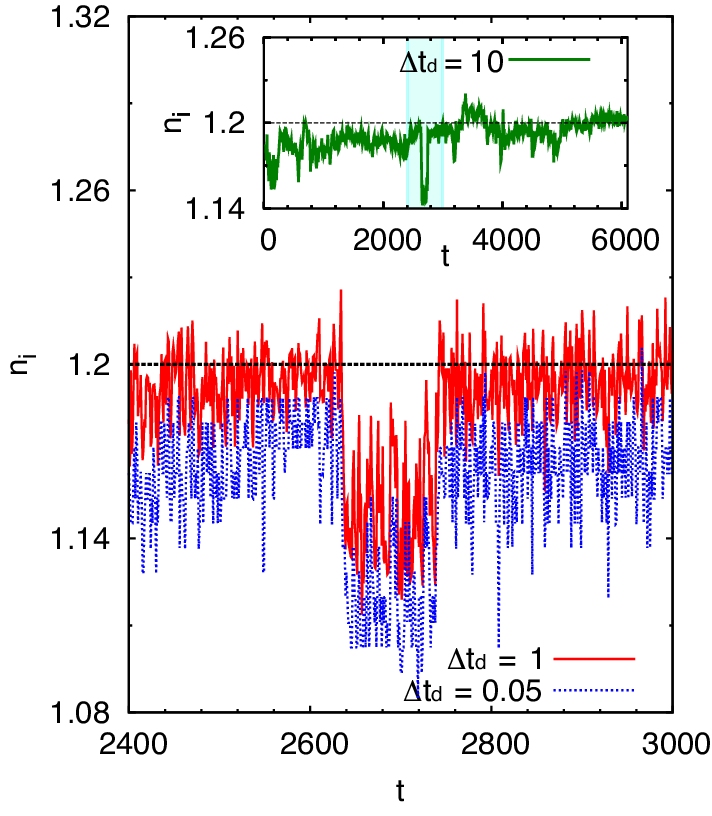}
\caption{\label{fig:mobile_free}
Comparison of $n_i(r_0)$ and its minimum per every unit time. 
The red solid line shows the local density $n_i (r_0)$ ($r_0=6.0\sigma_1$)
averaged over every unit time $(\Delta t_d=1)$ for a run at $T=0.56$. 
On the blue dashed line, $n_i(r_0)$ is calculated for
every time interval of $\Delta t_d=0.05$ and each minimum 
is plotted per unit time. Therefore, the latter dis systematically lower
than the former due to thermal fluctuations on a 
small time scale. Inset: For the same particle and for a longer time interval, 
$n_i(r_0)$ is plotted when averaged over a time interval of $\Delta t_d=10$.
}
\end{figure}

\begin{figure*}
\includegraphics[width=0.9\linewidth]{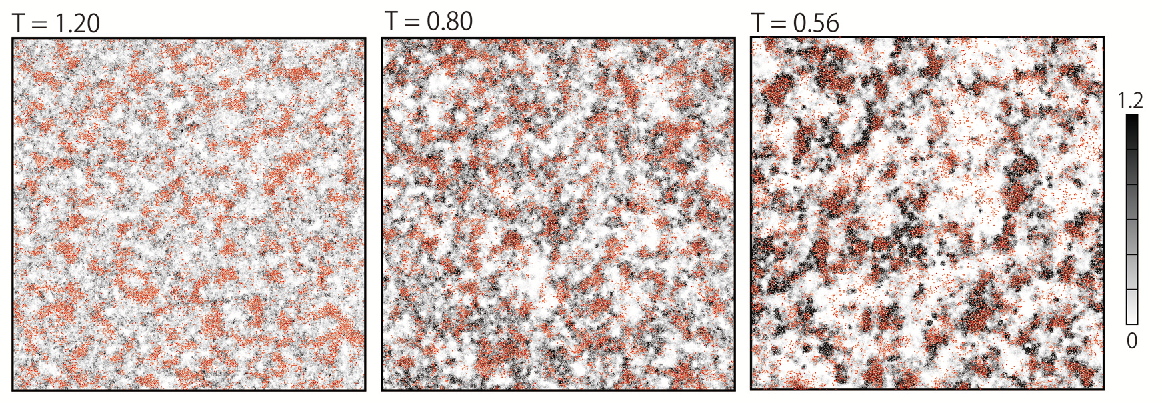}
\caption{\label{fig:freebb}
The red points show the distribution of 25\% of the particles having 
smaller values of minimum local densities $\langle \nu_i (\Delta t_b)\rangle_{\rm IE}$, 
in an IE
of 32 runs performed for $\Delta t_b = 6, 60,$ and $6100$ 
for $T=1.2, 0.80,$ and $0.56$, respectively. 
The number distribution of broken bonds
$\langle \mathcal{B}_i(\Delta t_b) \rangle_{\rm IE}$ 
is shown by the gray scale according to the scale bar on the right. 
}
\end{figure*}

\subsection{Broken bond and local free volume}
Dynamic heterogeneity is defined as a ubiquitous 
property of supercooled liquids in which configuration-rearranging regions 
emerge heterogeneously in the system. 
This property can be 
parametrized with various quantities 
including the van Hove correlation,\cite{97Kob} 
particle displacements,\cite{95Muranaka,Perera}
and four-point correlation functions. Here, to
characterize the mobile regions of dynamical heterogeneity,
we follow 
the method of broken bonds\cite{97YO_JPSJ,98YO_PRE,12SKO},
in which the number of broken-bond pairs for each particle 
is counted as $\mathcal{B}_i (\Delta t) = \sum_{j\in {\rm b.b.}} 1$.
Here, the summation is taken over $j$ 
particles at the broken bond ends of $i$ particles
calculated over time intervals of length $\Delta t$,
where the definition of 
broken bond pairs is given 
by Eqs. (\ref{eq:bonded}) and (\ref{eq:broken}). 
The broken bond distribution is a representation of irreversible 
configuration rearrangements, where collective motion due to 
long-wavelength sound modes is eliminated.  
To illustrate the irreversibility of the bond breakage, 
in Fig. \ref{fig:bond}, the total number of the
remaining bonds $N_b(t)$\cite{12SKO} is plotted for $T=0.64$, 
which is the number of unbroken bonds 
for the same thresholds $A_1$ and $A_2$.
Several pairs of particles $i,j$ satisfying Eq. (\ref{eq:bonded}) at time $t=0$
approach their counterparts again at time $t'$ after themselves undergoing breakage.
The number of these ``re-forming bonds'' is also shown in Fig. \ref{fig:bond},
where all the bond change is analyzed at every unit time. 
Because there are virtually no re-forming bonds at short time 
scales ($t\sim 10^2$), these bonds are not re-forming 
due to the vibration motion of the sound modes
but because of the successive particle rearrangements or 
cage jump events. 
A particle pair is likely to remain separated 
with a probability of 80\% once it becomes broken. 
As a result,  the time-scale of local diffusion around these breaking
bonds is characterized uniquely by bond-breakage relaxation 
time $\tau_b$.\cite{12SKO,13KO}
Thus, broken bond characterizes a more 
qualified aspect of dynamical heterogeneity. 

Since the spatial variation of dynamical heterogeneity strongly depends
on the particle positions, isoconfigurational ensembles (IEs)\cite{04WCHF,06WCHPRL} 
of 32 runs are employed and an average is calculated
over these runs.  
This IE average is denoted as $\langle \cdot\rangle_{\rm IE}$ in the following
discussions. In the top section of Fig. \ref{fig:FV_ST}, 
we plot the  $\left\langle \mathcal{B}_i (\Delta t) \right\rangle_{\rm IE}$ values for 
a short time interval, $t\in [0,\Delta t]\ (\Delta t=122)$, together with the distribution of 
2\% of particles having the lowest local densities $\langle n_i(r_0)\rangle_{\rm IE}$ 
taken at the initial stage $t\in [0,10]$. 
For $T=0.56$, this lapse of time $\Delta t = 122$ 
is taken at a bit longer time than the typical 
lifetime of the voids (see Fig. \ref{fig:mobile_free}), and thus, 
it corresponds to the time scale at  initial stage of
heterogeneous diffusion in the supercooled liquid.
The particles with low local densities can be divided into heterogeneous clusters by 
grouping particles separated by distances shorter than $9\sigma$ into one cluster.
These clusters are depicted by differently colored particles
in the bottom section of Fig. \ref{fig:FV_ST}. 
We can see that these clusters have length scales 
exceeding 10$\sigma$, 
which means that there are several numbers of 
free volumes within each cluster existing heterogeneously
in the supercooled state.  These clusters are spatially correlated 
with configuration rearrangements whose degree is 
represented by a variable 
$\langle \hat{\mathcal{B}} (\bm{r}, \Delta t)\rangle_{\rm IE}$, 
with $\hat{\mathcal{B}} (\bm{r}, \Delta t) = \sum_j \sigma_j^2 
\mathcal{B}_j (\Delta t) \delta (\bm{r}-\bm{r}_j (0) )$. \cite{12SKO}
In the bottom right part of Fig. \ref{fig:FV_ST}
which shows the average value 
$\langle \mathcal{B}_i\rangle_{\rm IE}$ for 
each of these clusters, we can see that all of these cluster
values largely exceed the total average value
$\mathcal{B}_0 = 0.0268$.  
It is also noteworthy that the larger clusters 
of free volumes shown in Fig. \ref{fig:FV_ST} as defined for
instantaneous time ($\Delta t = 10$) are 
spatially correlated well with long-time dynamics
as shown for $T=0.56$ in Fig. \ref{fig:freebb}.

To understand the cluster relationship with dynamical heterogeneity, 
it is necessary to understand how the clusters move over longer time intervals.
For $T=0.56, 0.64, 0.72, 0.80, 0.96, $ and $1.20$, 
simulation runs of IEs are performed 
with time intervals of $\Delta t_b (T) =$
6100, 815, 210, 80, 23, and 8, respectively.  
Upon using the concept of bond relaxation time $\tau_b (T)$ 
introduced in a previous study\cite{98YO_PRE} and defined by  
the relation 
$N_b (t_0 + \tau_b) = N_b (t_0)/e$,  these time intervals satisfy 
$\Delta t_b (T) \simeq 0.043\tau_b(T)$.
Here, $N_b(t)$ denotes the total number of 
initial bonds ($t=0$) remaining at time $t$,
and $\tau_b(T)$ characterizes the time scale of structural relaxation 
(see Table \ref{tab:relaxtime} for actual values).
In these intervals, almost identical portions of the
total initial bonds undergo breakage for all values of $T$.

For low $T$ values, wherein the system 
is dominated by slow dynamics,  
particle rearrangements agitated by thermal fluctuations
are expected to become intermittent. 
In Fig. \ref{fig:mobile_free}, for $T=0.56$, 
the change in the local density $n_i$ 
with time is measured for one specific particle in the
mobile region of dynamical heterogeneity. 
By taking the average of $n_i$ over every time interval of $\Delta t_d = 10$
as shown in the inset, the local density is observed
to fluctuate around $\phi$. 
In the region around $t=2700$,
the density reduces for a time interval of order $10^2$
by an amount of around $\Delta n_i \simeq 0.05$. Interestingly,
since $\Delta n_i\ \times\ \pi (r_0 /2)^2 \sim O(1)$, this reduction corresponds
to a free volume size of the scale of one particle, 
for the current radius $r_0=6.0$. 
The lower the temperature is, the longer the time interval is expected to
be due to smaller thermal fluctuations.
Upon decreasing the averaging time $\Delta t_d$ to 1 and 0.05, 
the temporal fluctuation of $n_i (r_0)$ appears more explicitly.
In the main graph, 
the red solid line represents $n_i (r_0)$ averaged over $\Delta t_d = 1$.
For a considerably shorter time interval of $\Delta t_d=0.05$, 
we accumulate the data of the density, and 
for every interval of unit time length ($\Delta t=1$), we take 
the minimum of $n_i(r_0)$. 
The blue dashed line represents the minimum value 
of the local density for each unit time thus taken.
Because these two lines 
follow parallel trajectories for low values of the density 
the minimum value in the local density history 
assumed by one particle can well represent and characterize
the general time development of the local density. 

\subsection{Heterogeneity  of local density}
Since the heterogeneity of free volumes shown in 
Fig. \ref{fig:FV_ST} is very weak, 
we define the ``minimum local density'' 
\begin{equation}
\nu_i (\Delta t) = \min_{t\in [t_0, t_1] } n_i,\quad t_1 = t_0+\Delta t.
\end{equation}
and study its heterogeneity. 
Because free volumes are of a size comparable to $\sigma_1$ 
lasting with a time scale longer than that of thermal fluctuations,
the IE average of 
$\hat{\nu}(\bm{r},\Delta t_b) = \sum_j \nu_j  (\Delta t_b) \delta (\bm{r}-\bm{r}_j(0) )$
can illustrate the spatiotemporal heterogeneity of free volumes in an 
enhanced manner.
Figure \ref{fig:freebb} shows the distribution of 25\% of the total number of particles
having lower $\langle \nu_i (\Delta t_b)\rangle_{\rm{IE}}$ values mapped together with
the broken bond distribution $\langle \mathcal{B}_i (\Delta t_b)\rangle_{\rm IE}$ 
The lower the value of $T$ is, the more distinct is the correspondence between
these distributions.  It is noteworthy that for $T=0.56$,
the instantaneous density distribution 
shown in Fig. \ref{fig:FV_ST} exhibits a correlation with that of the free volumes 
and the broken bonds shown in the figure to the right ($T=0.56$) in Fig. \ref{fig:freebb}.

\begin{figure}
\includegraphics[width=0.9\linewidth]{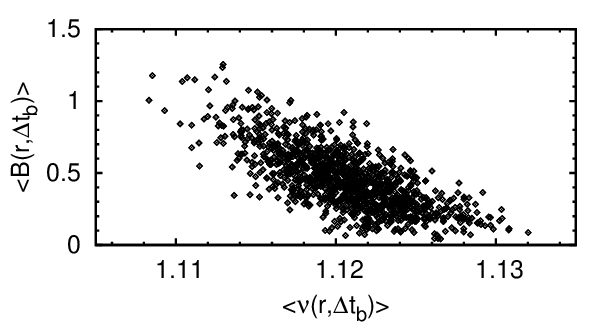}
\caption{\label{fig:scatter}
Scattered plot of cell-averaged quantities 
between $\langle {\hat{\nu}} (\bm{r},\Delta t_b)\rangle_{\rm IE}$ 
and $\langle \hat{\mathcal{B}} (\bm{r},\Delta t_b)\rangle_{\rm IE}$ for $T=0.56$.  
The cells divide the total system into a $12\times 12$ grid.
The data for runs with eight independent initial conditions are shown 
in this figure, and an IE average over 32 runs is taken for each run. 
}
\end{figure}

Quantitatively, this correspondence is illustrated for $T=0.56$ in Fig. \ref{fig:scatter};
the correspondence is observed as a scatter plot between
$\langle\hat{\nu} (\bm{r},\Delta t_b)\rangle_{\rm IE}$, 
with $\hat{\nu} (\bm{r},\Delta t_b) = \sum_j \nu_j \delta (\bm{r}-\bm{r}_j(0) )$
and $\langle\hat{\mathcal{B}} (\bm{r},\Delta t_b) \rangle_{\rm IE}$ averaged in each cell,
thereby squarely dividing the total system into a $12\ \times\ 12$ grid of 
size $(L/12)^2 = (23.4\sigma_1)^2$, 
which contains about 440 particles. 

\begin{figure}
\includegraphics[width=0.88\linewidth]{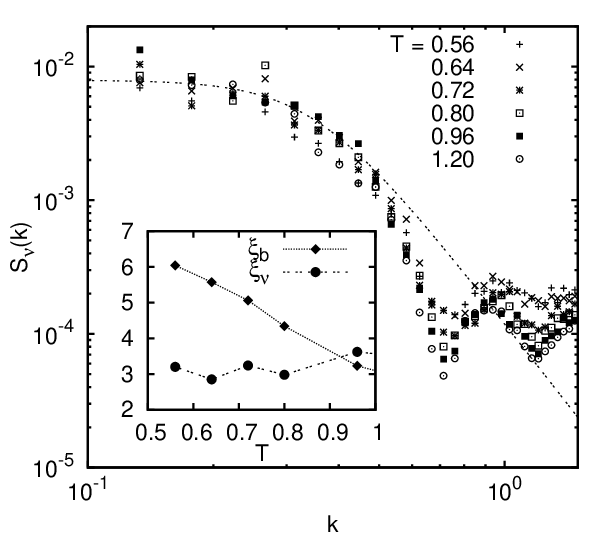}
\caption{\label{fig:STRAVE}
Structure factor of the minimum local density
 $S_\nu (k, \Delta t_b)$
defined in Eq. (\ref{eq:STRAVE}). 
For various values of $T$, the data are well
fitted by $f(k)= C_0 (1+k^4\xi_\nu^4)^{-1}$.
The dotted line shows the fit at $T=0.72$. 
Inset: Correlation lengths of broken bond distribution $\xi_b$
and those of ``minimum local density'' $\xi_\nu$ for various values of $T$. 
While the former is estimated by assuming 
the Ornstein-Zernike form ($\sim (1+k^2\xi_b^2)^{-1}$ ), 
the latter is plotted by using a sharper form 
decaying as $k^{-4}$ at $k\sim 0.5$. 
}
\end{figure}

To examine how the heterogeneity in the density evolves, 
in Fig. \ref{fig:STRAVE}, 
we show the structure factor of 
$\hat{\nu} (\bm{r},\Delta t_b)$
defined as 
\begin{equation}
S_\nu (k,\Delta t_b) = \langle |\hat{\nu}_{\bm{k}} (\Delta t_b)|^2\rangle_{\rm IE},
\label{eq:STRAVE}
\end{equation}
where $\delta\hat{\nu} (\bm{r}, \Delta t_b) = \sum_j \sigma_j^2 (\nu_j - [\nu_j ]_N ) \delta (\bm{r}-\bm{r}_j (0) )$
denotes the mesoscopic fluctuation of $\nu_i$, and $\hat{\nu}_{\bm{k}} (\Delta t_b)$ indicates 
the Fourier components of $\delta\hat{\nu} (\bm{r},\Delta t_b)$. 
The average is taken over IEs of 128 (32 for $T\le 0.64$) runs
generated with 32 independent initial configurations, and thus,
data from 4096 (or 1028) runs performed over a duration of $\Delta t_b$ 
are used for each $T$. 
Since one isolated free volume lowers the values of $n_i(r_0)$ of particles 
within the distance of $l = r_0\sigma_1$,
we observe a small peak in $S_\nu (k,\Delta t_b)$ around  $k \sim 2\pi / l \simeq 10^0$.
Remarkably, 
$S_\nu (k,\Delta t_b)$ 
is enhanced at low-$k$, 
exhibiting the same degree of enhanced heterogeneity.
Because the shape of each lower-density region illustrated in Fig. \ref{fig:freebb}
shows sharp boundaries, it is much better fitted with 
$f(k) = C_0 (1+k^4\xi_\nu^4 )^{-1}$ than the Ornstein-Zernike (OZ) form. 
In the inset of Fig. \ref{fig:STRAVE}, we compare the correlation lengths $\xi_\nu$ and $\xi_b$ 
of the minimum local density and the broken bond distributions,
estimated by fitting $f(k)$ to $S_\nu (k,\Delta t_b)$ 
and the OZ form to $S_b (k,\Delta t_b)$ (see Refs. \onlinecite{97YO_JPSJ,98YO_PRE}), 
respectively. 
For the interval of the corresponding structural relaxation time $\Delta t_b(T)$
the heterogeneity in $\hat{\nu}(\bm{r},\Delta t_b)$  is nearly 
independent of the  temperature, 
while $S_b(k,\Delta t_b)$ displays
the OZ form with growing length scales for 
lower values of $T$.

\subsection{Normal mode analysis}
Recent numerical simulations evidence that 
the static particle configuration itself determines the dynamic propensity
distribution, even in an amorphous state apparently lacking 
structural order.\cite{04WCHF,06WCHPRL}
In particular, localized soft modes predicted by the
``normal mode analysis'' have been revealed to be a good predictor
of the dynamic heterogeneity; from a single and instantaneous snapshot
of the particle configuration, information regarding the spatial heterogeneity 
of dynamics can be extracted with the use of the 
distribution of localized low-frequency phonons.\cite{08WC, 12Matsuoka}
Similar methods are also employed to 
predict yielding spots in a jammed system
at zero temperature.\cite{10Chen,11Manning} 

To compare the distribution of the free volumes 
with the high-propensity region determined from a static configuration,
we perform the normal mode analysis for our system
along the lines of Ref. \onlinecite{08WC}.
After the configuration of the particles $\{\bm{r}_i \}$ is quenched
to a state with local stability with the use of the conjugate gradient method, 
the solution $\omega=\omega_k$ of the eigenvalue equation 
$({\sf D}_{i\alpha, j\beta} -\omega^2 ) a_{j\beta} = 0\ (\alpha,\beta \in \{x,y\}$)
for the $2N\times 2N$ Hessian matrix
\begin{equation}
{\sf D}_{i\alpha, j\beta} = \frac{1}{\sqrt{m_im_j}} \left(\frac{\partial^2 V(\bm{r}) }{\partial\bm{r}_i^\alpha \partial\bm{r}_j^\beta} \right), 
\end{equation}
is calculated, where $i$ and $j$ denote particle indices.
The eigenvalues $\omega_k$ and the eigenvectors $\vec{a}_{j\beta}=a_{j\beta,k}$ 
represent the frequencies of the normal modes 
and the amplitudes for $k$-th mode respectively, the $1\le k\le 2N$ being the mode index. 
Excluding the two lowest-frequency modes corresponding to the translational motion, 
we consider the thermal vibration amplitude, {\it i.e.} the local Debye-Waller factor 
\begin{equation}
\mathcal{W}_i = \frac{k_BT}{m_i} \sum_{k=2}^M  \frac{ a_{ix,k}^2 + a_{iy,k}^2 }{\omega_k^2}, \label{eq:dwf}
\end{equation} 
where $M$ denotes the
 number of the lowest-frequency modes taken into account. 
This quantity represents the strength of the squared vibration amplitude
for each particle for small time scales, 
{\it i.e.}, localized long-wavelength soft modes represent 
the propensity distribution.
Upon using the equipartition rule, the thermal energy for each mode is 
given by $k_BT$,  which  means that
we should weight it by inverse-square frequencies in the superposition of the 
non-dimensional amplitudes $A_{j\beta,k}$.

\begin{figure}
\includegraphics[width=0.95\linewidth]{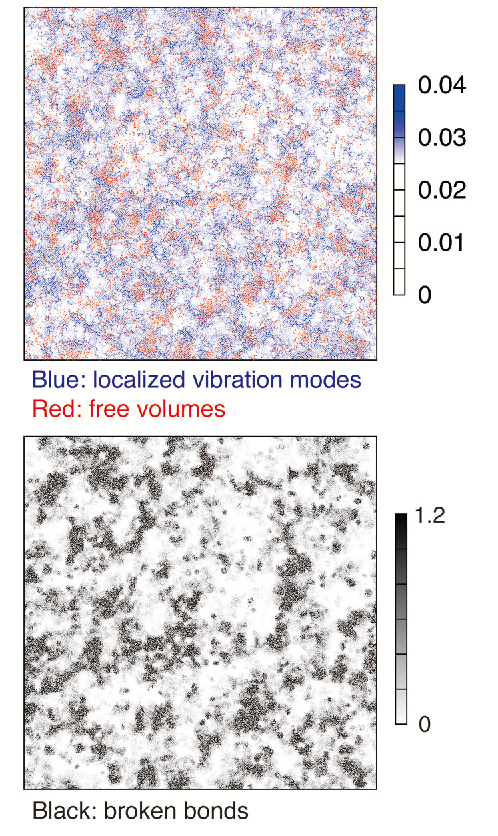}
\caption{\label{fig:NORMAL}
Top: The local Debye-Waller factor in Eq. (\ref{eq:dwf}),
where the number of the modes $M$ is 2560, is shown 
with the blue-colored map for $T=0.56$ for the same snapshot
as in Fig. \ref{fig:freebb}, together with the 
25\% of the particles having 
smaller values of $\langle \nu_i (\Delta t_b)\rangle_{\rm IE}$,
{\it i.e.}, $\langle \nu_i (\Delta t_b)\rangle_{\rm IE} < 1.11107$. 
Bottom: The corresponding map of 
the  number distribution of broken bonds for the same IE 
$\langle \hat{\mathcal{B}}(\bm{r},\Delta t)\rangle_{\rm IE}$ 
for $T=0.56$ as in Fig. \ref{fig:freebb}.
}
\end{figure}

In Fig. \ref{fig:NORMAL}, the spatial distribution of 
the localized soft modes for $M=2560$ 
is shown with the blue-colored map
in the top panel, wherein the red particles represent 
the free volume distribution (the same as 
those for $T=0.56$ in Fig. \ref{fig:freebb}).
Both the distributions have 
substantial overlaps with that of broken bonds, as shown
separately in the bottom panel of Fig. \ref{fig:NORMAL}.
Free volumes are relatively absent around 
particles with smaller vibration amplitudes, as
represented by blanks in the blue-colored plot. 
We can also confirm the tendency
by measuring the average values over
a $24\times 24$ divided mesh, each of which contains about 110 particles.
None of the 71 cells having small values of the Debye-Waller factor
($[{\mathcal{W}}]_\Sigma < 0.027$) has the averaged local minimum density
$\langle \nu_i (\Delta t_b)\rangle_{\rm IE} > \nu_0 = 1.11107$,
where $[\cdot ]_\Sigma$ denotes the average over a cell. 

\begin{figure}
\includegraphics[width=0.95\linewidth]{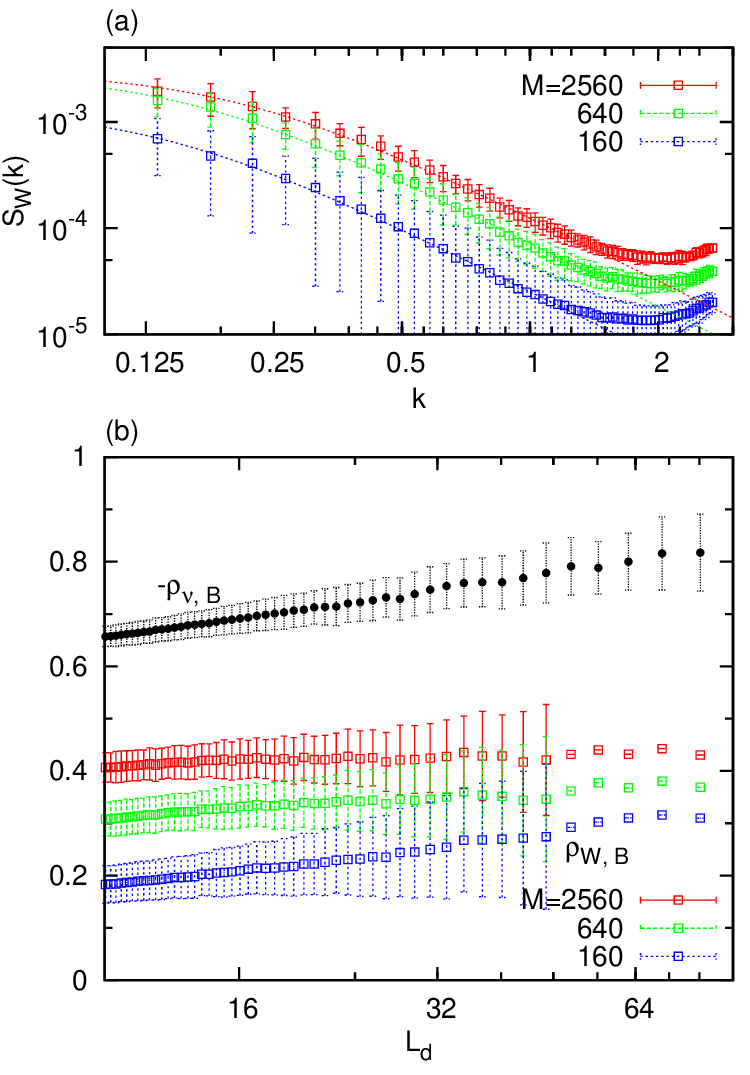}
\caption{\label{fig:corr}
(a) The structure factor $S_{\mathcal{W}}$
of the thermal vibration amplitude is shown for 
$M=$\ 160, 640, and 2560 lowest eigen-frequencies.
Dotted lines represent the fit by the OZ form 
$f(k)\sim (1+k^2\xi_M^2)^2$ for each $M$.
(b) The Pearson correlation coefficients $\rho_{\mathcal{W,B}}$ for 
$M=$160, 640, and 2560, and also the minus of $\rho_{\nu, \mathcal{B}}$
are shown for various box cell sizes $L_d$.
The data represent the degree of spatial correlation between
$\mathcal{W}_i$ and $\mathcal{B}_i$. 
The error bars are abbreviated  at large $L_d$ values 
for the sake of visibility.
}
\end{figure}

While the spacial distribution of the superposition 
${\mathcal W}_i$ exhibits localization of the vibration modes, 
a large-scale heterogeneity can be observed,  
as observed from Fig. \ref{fig:NORMAL}. In Fig. \ref{fig:corr} (a), 
the structure factor of $\mathcal{W}_i$ defined by
\begin{equation}
S_{\mathcal{W}}(k) = \langle |\hat{ \mathcal{W} }_{\bm{k}} |^2\rangle ,
\end{equation}
where $\hat{\mathcal{W}} = \sum_j \sigma_j^2  \mathcal{W}_i
\delta (\bm{r}-\bm{r}_j(0) )$ 
is shown for various numbers of the 
lowest-frequency modes $M=160,\ 640,$ and 2560. 
These data can be fitted well to the OZ form $f(k) = (1+k^2\xi_M^2)^{-1}$, 
where the correlation lengths $\xi_M$  are estimated to be 
$\xi_M \simeq 7.3\pm0.3,\ 6.3\pm 0.3,$ and $4.8\pm 0.2$, respectively. 
On the one hand $S_{\mathcal{W}}(k)$ continues to increase for
low values of $k$ for small $M$(=160) because 
these low normal modes 
have large scale characteristic lengths, 
but at the same time it stops growing at low $k$ 
for $M=2560$ due to the localization.  

To investigate the length scales involved in the 
correspondence between the heterogeneities 
presented above, we calculated the 
coefficients of the spatial correlations of these quantities 
in the following manner:
the system is divided into boxes with lengths $L_d = L/d\ (d = 2,3,\cdots)$. 
Within each of these boxes, we estimate the average values 
$[\nu]_\Sigma,\ [\mathcal{B}]_\Sigma,\ $ and $[\mathcal{W}]_\Sigma$,
and subsequently, we calculate $\rho_{\mathcal{W,B}}$ and $\rho_{\nu, \mathcal{B}}$
over $d^2$ data points, where $\rho_{\alpha,\beta}$ denotes
the Pearson correlation coefficients for a pair of statistical data series
(denoted by $\alpha$ and $\beta$) defined by
\begin{equation}
\rho_{\mathcal{\alpha,\beta}} = \frac{\sum \left( \alpha - E( \alpha) \right) 
\cdot \sum ( \beta - E(\beta) ) }{ \left[\sum (\alpha - E(\alpha))^2 \right
]^{1/2} \left[ \sum (\beta - E(\beta) )^2 \right]^{1/2}}, 
\end{equation}
where $\{\alpha,\beta\} \in \{ [\nu]_\Sigma,\ [\mathcal{B}]_\Sigma,\ [\mathcal{W}]_\Sigma\}$
and $E(X)$ denotes the average of $X$ over all the boxes.
This quantity assumes a value of 1 (or -1) when 
$\alpha$ and $\beta$ are perfectly correlated (anticorrelated), 
and becomes 0 if there are no correlations.
In Fig. \ref{fig:corr}, 
together with the 
negative value of $\rho_{\nu, \mathcal{B}}$
representing the anticorrelation between the local density and the broken bonds, 
$\rho_{\mathcal{W,B}}$  for $M=$\ 160, 640, and 2560
is plotted at various cell sizes $L_d$ 
whose  value corresponds to the degree of coarse-graining. 
The average is taken over calculations for 16 independent initial configurations. 
Since the vibrational amplitude $\mathcal{W}_i$ is more 
directly linked with the short-time vibration motion rather than the 
broken bond distribution, it has smaller values
and larger statistical errors than $-\rho_{\nu, \mathcal{B}}$,
but it assumes 
definitely positive correlations. 
A noteworthy observation is that, 
at a large scale ($L_d \sim 100$),
the correlation coefficient of $\mathcal{W}_i$ for $M=160$ 
with  $\mathcal{B}_i$ is as large as that for $M=2560$. 
This suggests that the nature of the occurrence of 
long-range heterogeneity in the dynamics 
can more or less be described by the long-ranged 
low-frequency sounds modes, and that
the high-frequency modes are relatively 
irrelevant to the long-ranged dynamical heterogeneity.

\section{Conclusion} \label{sec:conclusion}
In this study, the relationship between mesoscopic heterogeneities of 
free-volume distribution, soft-mode localization, and broken bonds has been investigated.  
The results show that they are largely correlated with a clear overlap.
However, the correlation lengths have qualitatively different 
characteristics--extensive studies on this topic have revealed that
dynamical heterogeneity in a supercooled liquid accompanies largely growing correlation 
lengths in the dynamics, 
as is also shown in this study in terms of 
the parameter $\xi_b$ in Fig. \ref{fig:STRAVE}.
In contrast, the correlation length of the free volume distribution
represented by the ``minimum local density'' does not exhibit growth
and remains short-ranged even at a low temperature. 
Thus, free volumes are distributed over a small range in the regions where 
configuration rearrangements occur as seeds for dynamic heterogeneity 
at large lengths scales.
By performing the normal mode analysis, 
we find  that the lowest-frequency $M=160$ 
sound modes, which form about $0.1\%$ of 
the total vibration modes, 
describes the most of the relationships between
the localized soft modes (for $M=2560$) and the 
dynamical heterogeneity.
Although there is a missing link between the 
vibrational spectra and broken bond distributions, 
the data suggest the possibility that
large-scale correlation originates from interactions
between fragile regions facilitated by large
scale sound modes stemming over these regions,
and that the free volume is a candidate for the representation
of these fragile regions.

Recently, experiments on colloidal clusters\cite{11Yunker,13Yunker}
and simulations\cite{13KawasakiJCP} reveal a
correlation between the neighbor number, 
vibration mode, and irreversible rearrangements. 
The correlation between the localized vibration modes and 
the free volumes is attributed to relatively smaller 
neighbor numbers around the local free volumes, 
which we speculate as providing collateral evidence for 
the role of free volume in the heterogeneity. 
However, the nature of the causal mechanism because of 
which free volume distributions bring about the long-wavelength 
sound vibration is still an open question,

We add some remarks for clarification in the following: 

(i) The fact that the minimum local density heterogeneity does not 
depend on the temperature for the same degree of frustration 
indicates that what we observe as density heterogeneity 
is not specific to the supercooled state but is inherently present 
even in simple liquids. Because the configuration changes becomes slower
as the system approaches the glass transition, the cooperative 
motion of free volumes increases at a longer-range  to enhance dynamical heterogeneity.

(ii) While in a number of experiments
long-ranged correlations of the density in glass have been observed,\cite{91Fischer,91vanMegen,93Fischer,94Kanaya,00Fischer}
in the simulation of models with repulsive cores presented in this study, 
the bare structure factor $S(k)$ does not exhibit growth at  long wavelengths, similar to the results of other such 
studies.\cite{98YO_PRE,10TanakaNMat} Though the possibility of longer-range density correlation
is not excluded, it would be weak in a model with a strong frustration.
 It is also noteworthy that 
in a single component 2D crystal of Lennard-Jones system, 
where density fluctuation is enhanced at long-wavelengths,  
the long-ranged dynamic heterogeneity is brought 
about by a local diffusion of defects with lower local densities rather
than by a long-ranged static fluctuation,\cite{09ShibaEPL}
in analogy with our current simulation.

(iii) Instantaneous normal modes have been analyzed at 
a low temperature ($T=0.56$) as having a large correlation length.
No related studies on the analysis of such modes 
for a system size larger than that 
used in this study are available; in our study,  
eigenvectors of a $128000^2$ matrix have been calculated. 
The analysis of localized soft modes indicates that 
a small number of lowest-frequency soft modes extend over large length 
scales, endowing the system with long-ranged heterogeneity. 
In the same model system, large-scale vibration motions
throughout the whole system 
have been observed to be enhanced because 
the system becomes
more and more rigid as we lower the temperature.\cite{12SKO}
These vibrations may possibly 
facilitate interactions between the far-distance fragile regions, 
to result in long-range dynamical critical fluctuations. 
Further studies are necessary to reveal the interplay between
the sound modes and the configuration rearrangement.

(iv) Although we have limited ourselves to investigation on 
the potential role of density fluctuations, 
it is still an open question what the primary causal 
reason for the dynamical heterogeneity.
The role of other possible static origins
that can affect the dynamics,  
for example, density gradient, bond orientation order, and so on, 
should be investigate further, which is beyond the 
current scope of the paper.

\begin{acknowledgements}
The authors thank A. Ikeda, K. Miyazaki, A. Onuki, K. Kim, H. Mizuno, Y. Noguchi,
and P. Harrowell for enlightening discussions.
The numerical calculations were carried out on the SGI Altix ICE 8400EX
and NEC SX9 systems at ISSP, University of Tokyo.
This work is supported by the Core-to-Core Program 
``International research network for non-equilibrium dynamics of soft matter''
by the Japan Society for Promotion of Science (JSPS),  
and also partially by a Grant-in-Aid for Scientific Research on Innovative Areas
``Synergy of Fluctuation and Structure: Foundation of Universal Laws in
Nonequilibrium Systems'' (Grant No. 25103010).
T. K. was supported by JSPS Research Fellowships for Young Scientists (Grant No. 10J02221).
\end{acknowledgements}

\end{document}